\documentclass[12pt]{article}
\usepackage{fullpage}
\usepackage{float}
\usepackage{floatflt}
\usepackage[centertags]{amsmath}
\allowdisplaybreaks[1]
\usepackage{amsbsy}
\usepackage{amsfonts}
\usepackage{amssymb}
\usepackage[dvips]{graphicx}
\usepackage{url}
\begin{document}

\begin{flushright}
\textsf{28 May 2003}
\\
\textsf{physics/0305122}
\end{flushright}

\vspace{1cm}

\begin{center}
\large
\textbf{Lorentz Invariance of Neutrino Oscillations}
\normalsize
\\[0.5cm]
\large
C. Giunti
\normalsize
\\[0.5cm]
INFN, Sezione di Torino, and Dipartimento di Fisica Teorica,
\\
Universit\`a di Torino,
Via P. Giuria 1, I--10125 Torino, Italy
\\[0.5cm]
\begin{minipage}[t]{0.8\textwidth}
\begin{center}
\textbf{Abstract}
\end{center}
\small
It is shown that,
in spite of the appearances, 
the standard expression for the
oscillation probability of ultrarelativistic neutrinos
is Lorentz invariant.
\end{minipage}
\end{center}

Flavor is the quantum number that distinguishes the different types
of quarks and leptons.
It is a Lorentz-invariant quantity.
For example,
an electron is seen as an electron by any observer,
never as a muon.
Therefore,
the probability of flavor neutrino oscillations
must be Lorentz invariant
(as already remarked in Ref.~\cite{Giunti:2000kw}).
However,
it may appear that the standard expression for the
probability of $\nu_\alpha\to\nu_\beta$
transitions
in vacuum\footnote{
In the simplest case of two-neutrino mixing with
$ \nu_\alpha = \cos\vartheta \nu_1 + \sin\vartheta\nu_2 $
and
$ \nu_\beta = -\sin\vartheta \nu_1 + \cos\vartheta\nu_2 $,
the transition probability in Eq.~(\ref{001}) reduces to the well
known expression
$
P_{\nu_\alpha\to\nu_\beta}(E,L)
=
\sin^2 2\vartheta
\,
\sin^2\left( \frac{ \Delta{m}^2 L}{ 4 E } \right)
$.
},
\begin{equation}
P_{\nu_\alpha\to\nu_\beta}(E,L)
=
\sum_k |U_{\alpha k}|^2  |U_{\beta k}|^2
+
2
\mathrm{Re}
\sum_{k>j}
U_{\alpha k}^* U_{\beta k} U_{\alpha j} U_{\beta j}^*
\,
e^{-i\Delta\phi_{kj}(E,L)}
\,,
\label{001}
\end{equation}
with the phase differences
\begin{equation}
\Delta\phi_{kj}(E,L)
=
\frac{ \Delta{m}^2_{kj} \, L }{ 2 \, E }
\,,
\label{008}
\end{equation}
is not Lorentz invariant.
In Eq.~(\ref{001}) $U$ is the mixing matrix that connects
the flavor neutrino fields
$\nu_{\alpha}$ ($\alpha=e,\mu,\tau$)
with the massive neutrino fields
$\nu_k$ ($k=1,2,3$)\footnote{
If the number of massive neutrinos is $N>3$,
the indices $k,j$ run from $1$ to $N$
and $\alpha,\beta=e,\mu,\tau,s_1,\ldots,s_{N-3}$,
with the indices
$s_1,\ldots,s_{N-3}$
denoting $N-3$ sterile neutrino fields
(see Refs.~\cite{BGG-review-98,Gonzalez-Garcia:2002dz}).
},
$L$ is the distance between the neutrino source and the neutrino detector,
and $E$ is the neutrino energy
(see Refs.~\cite{BGG-review-98,Gonzalez-Garcia:2002dz}).
In Eq.~(\ref{008})
$\Delta{m}^2_{kj} \equiv m_k^2 - m_j^2$
is the difference between the squared-masses of $\nu_k$ and $\nu_j$.

Let us assume that Eq.~(\ref{001}) is valid in the inertial system $\mathcal{O}$
with time axis $t$ and the $x$ axis in the direction of neutrino propagation.
Consider another inertial system
$\mathcal{O}'$ with axes $x',t'$
moving with respect to
$\mathcal{O}$ with velocity $v$ in the $x$ direction.
The Lorentz transformations of space and time intervals are\footnote{
We use natural units, with $c=1$.
}
\begin{subequations}
\label{002}
\begin{align}
\null & \null
\Delta{x}' = \gamma \left( \Delta{x} - v \, \Delta{t} \right)
\,,
\label{002a}
\\
\null & \null
\Delta{t}' = \gamma \left( - v \, \Delta{x} + \Delta{t} \right)
\,,
\label{002b}
\end{align}
\end{subequations}
with
$ \gamma \equiv \left( 1 - v^2 \right)^{-1/2} $.
The length
$\Delta{x}'$
in the system $\mathcal{O}'$
of an object with length $\Delta{x}$
in the system $\mathcal{O}$
is calculated at equal times in the system $\mathcal{O}'$:
\begin{equation}
\Delta{t}' = 0
\quad
\Longrightarrow
\quad
\Delta{t} = v \, \Delta{x}
\quad
\Longrightarrow
\quad
\Delta{x}' = \gamma \left( 1 - v^2 \right) \Delta{x} = \frac{ \Delta{x} }{ \gamma }
\,.
\label{003}
\end{equation}
This is the well-known Lorentz contraction of distances.
The Lorentz-contracted distance between neutrino source and detector
measured in the system $\mathcal{O}'$ is given by
\begin{equation}
L' = \frac{ L }{ \gamma }
\,.
\label{004}
\end{equation}
On the other hand,
the Lorentz transformations of momentum and energy are
\begin{subequations}
\label{005}
\begin{align}
\null & \null
p' = \gamma \left( p - v \, E \right)
\,,
\label{005a}
\\
\null & \null
E' = \gamma \left( - v \, p + E \right)
\,.
\label{005b}
\end{align}
\end{subequations}
In the massless limit
\begin{equation}
E = p
\quad
\Longrightarrow
\quad
E' = p' = \gamma \left( 1 - v \right) E
\,.
\label{006}
\end{equation}
It is clear that if the values of $L'$ and $E'$
in Eqs.~(\ref{004}) and (\ref{006})
were correct for the calculation of the oscillation probability
in the system
$\mathcal{O}'$,
the resulting oscillation probability
$P_{\nu_\alpha\to\nu_\beta}(E',L')$
would be different from
$P_{\nu_\alpha\to\nu_\beta}(E,L)$,
leading to an absurd violation of the Lorentz invariance of flavor.

A possible objection to this reasoning is that the phase differences in Eq.~(\ref{008})
are valid only in the reference frame in which the
source and detector are at rest.
The answer to this objection is that
the expression in Eq.~(\ref{008}) is derived
without any assumption of a special reference frame,
and hence it must be valid in any frame.
Indeed,
the phase differences in Eq.~(\ref{008})
follow from a simple relativistic approximation of the explicitly
Lorentz invariant phases
\begin{equation}
\phi_k(T,L) = E_k \, T - p_k \, L
\,,
\label{007}
\end{equation}
where $T$ is the time of propagation of the neutrino
from the source to the detector,
$E_k$ and $p_k$
are the energy and momentum of $\nu_k$.
For $T=L$, corresponding to the leading order contribution
to the phase of ultrarelativistic neutrinos
(see Ref.~\cite{Giunti:2000kw}),
we have
\begin{equation}
\phi_k(T=L)
=
\left( E_k - p_k \right) L
=
\frac{ E_k^2 - p_k^2 }{ E_k + p_k } \, L
=
\frac{ m_k^2 }{ E_k + p_k } \, L
\simeq
\frac{ m_k^2 }{ 2 \, E } \, L
\,,
\label{0071}
\end{equation}
where we have used the relativistic dispersion relation
$ E_k^2 = p_k^2 + m_k^2$,
and
$E$ is the neutrino energy in the limit $ m_k \to 0 $.
The difference of the phases in Eq.~(\ref{0071}) is the phase difference in Eq.~(\ref{008}).
From this simple derivation in which we did not make any assumption
on the reference frame,
it is clear that the phases (\ref{0071}) and the phase differences (\ref{008})
are valid in any frame and should be Lorentz invariant as the phase
(\ref{007})
from which they are derived.

The above derivation helps in understanding the 
solution of the problem:
the crucial point is the approximation
$T=L$,
which means that $L$ is not the equal-time distance between source and detector,
but the distance traveled by the neutrino in the time $T$.
In other words,
in any system $L$ is the spatial distance between the space-time
events of neutrino production and detection.

If in the system $\mathcal{O}$ we have
$\Delta{x}=\Delta{t}$,
from Eqs.(\ref{002}) in the system $\mathcal{O}'$ we have
\begin{equation}
\Delta{x}' = \Delta{t}' = \gamma \left( 1 - v \right) \Delta{x}
\,.
\label{009}
\end{equation}
Therefore, the correct transformation for the distance in the phases is not
given by the Lorentz contraction formula in Eq.~(\ref{004}),
but by
\begin{equation}
L' = \gamma \left( 1 - v \right) L
\,.
\label{010}
\end{equation}
Confronting this expression with the transformation of energy in Eq.~(\ref{006}),
it is clear that the ratio
$L/E$
is Lorentz invariant,
as well as
the phases in Eqs.~(\ref{0071}) and (\ref{008}),
and the flavor transition probability in Eq.~(\ref{001}).

A simple example can show clearly
that the Lorentz-contracted source-detector distance (\ref{004})
calculated at equal times in the system $\mathcal{O}'$
does not correspond to the distance traveled by the neutrino.
Consider the source and detector at rest in the system $\mathcal{O}$.
Then in the system $\mathcal{O}'$
the detector is moving with velocity $-v$ along the $x$ axis.
Since the detector moves
after the propagating neutrino has left the source,
the spatial distance traveled by the neutrino
is shorter than the instantaneous source-detector distance.

\begin{figure}[p]
\begin{center}
\includegraphics*[bb=82 270 543 729, width=\textwidth]{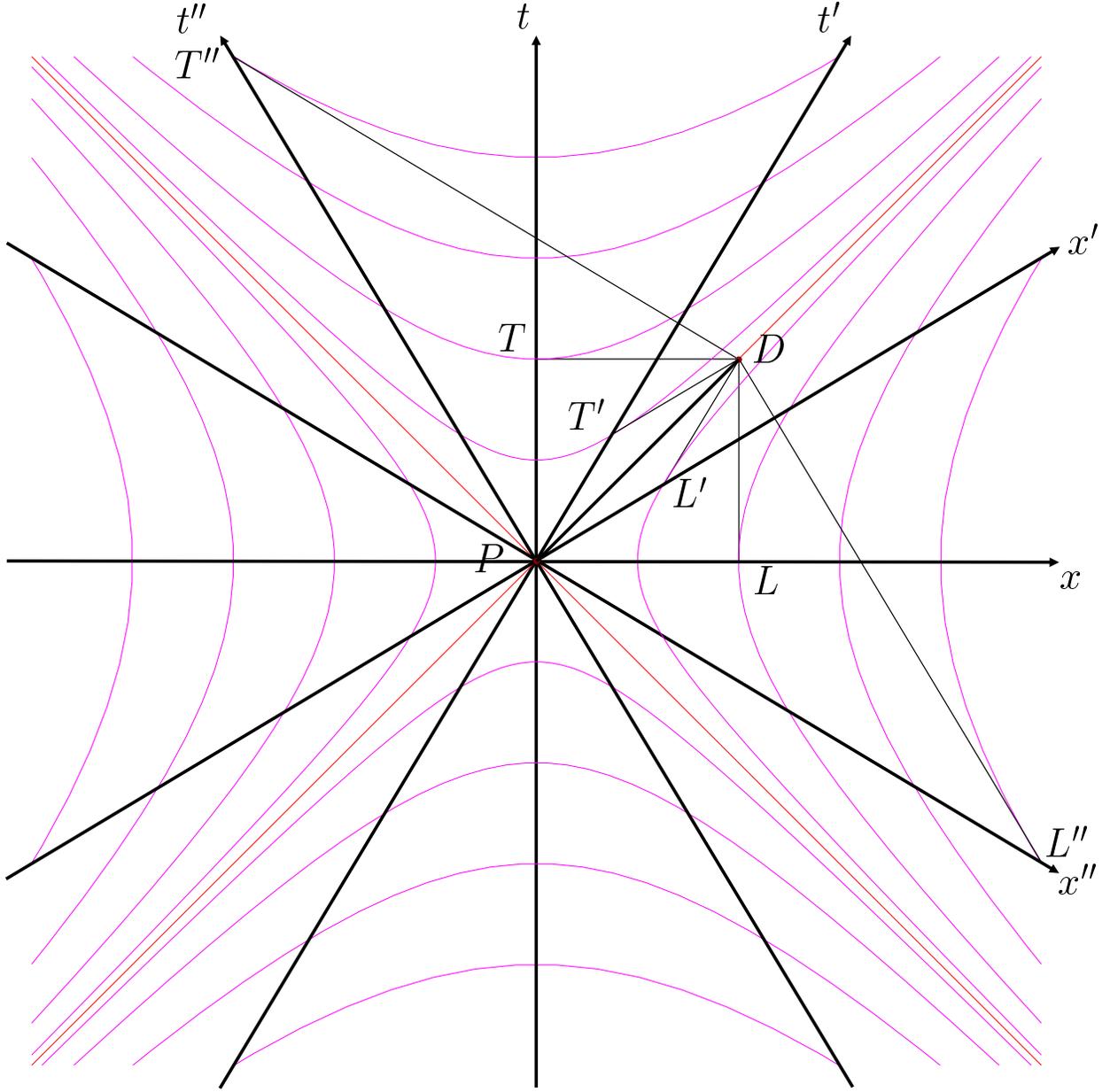}
\end{center}
\caption{ \label{min}
Minkowski diagram showing the distance and time between the space-time events of
neutrino production $P$
and detection $D$
measured in three coordinate systems:
$\mathcal{O}$ with axes $x,t$;
$\mathcal{O}'$ with axes $x',t'$
moving with respect to
$\mathcal{O}$ with velocity $v=3/5$ in the $x$ direction along neutrino propagation;
$\mathcal{O}''$ with axes $x'',t''$
moving with respect to
$\mathcal{O}$ with velocity $v=-3/5$ in the $x$ direction.
The $x'$ ($x''$) axis is inclined with respect to the $x$ axis
by an angle $\arctan 3/5$ ($-\arctan 3/5$);
the $t'$ ($t''$) axis is inclined with respect to the $t$ axis
by an angle $-\arctan 3/5$ ($\arctan 3/5$).
The hyperbolas
$t^2-x^2={t'}^2-{x'}^2={t''}^2-{x''}^2=\mathrm{constant}$
fix the scale on the axes.
One can see that
$L'=L/2$ and $L''=2L$,
in perfect agreement with Eq.~(\ref{010}).
}
\end{figure}

The correct transformation law of the propagation distance in Eq.~(\ref{010})
is illustrated by the Minkowski diagram
in Fig.~\ref{min},
in which one can clearly see that in a system in which source and detector are in motion
the spatial distance
between the two space-time events of neutrino production and detection is different
from the instantaneous source-detector distance.
The two distances coincide only in the system $\mathcal{O}$,
in which source and detector are at rest.

Finally,
one can ask what happens if instead of considering the source and detector at rest in
the system $\mathcal{O}$,
we consider a case in which source and detector are in relative motion.
Since the oscillation probability is measured by the detector,
the velocity of the source with respect to the detector
does not matter\footnote{
Of course, the velocity of the source
must be taken into account in the calculation of
the neutrino energy.
For example,
if the source is a decaying pion,
the neutrino energy depends on the velocity of the pion.
}.
The system $\mathcal{O}$
in which the distance $L$ covered by the neutrino
coincides with the instantaneous distance between source and detector
at the time of neutrino emission
is always the rest system of the detector.
In Fig.~\ref{min} the world-line of the detector is constrained to be the vertical line passing
through the space-time event $D$ of neutrino detection,
but the world-line of the source
can be any time-like line passing through the space-time event $P$ of neutrino production.

\section*{Acknowledgments}
\label{Acknowledgements}

I would like to thank the
School of Natural Sciences,
Institute for Advanced Study,
Princeton
for kind hospitality during the completion of this work.

\end{document}